\documentstyle[aps]{revtex}

\begin{document}
\draft
%\preprint{Preprint }
\title{Comment on "Transverse Force on a Quantized Vortex
in a Superfluid" }
\author{  G.E. Volovik}

\address{Low Temperature Laboratory, Helsinki University of
Technology, 02150 Espoo, Finland, \\ and \\ Landau Instute for
Theoretical
Physics,
117334
Moscow, Russia}

\date{\today} \maketitle

%\eject
\narrowtext
\twocolumn

In a recent Letter \cite{Thouless} Thouless {\sl et al} (TAN)
suggested an exact expression for the nondissipative transverse force
on a vortex line and claimed that  it contained no
contribution from fermions localized in the vortex core.

The forces on the vortex have been recently measured in superfluid
$^3$He-B in  a broad temperature range  \cite{Bevan1995}. A general
expression for
a balance of forces acting on the vortex with circulation
$\kappa$ moving with
velocity
${\bf v}_V$ is
\cite{Sonin}
\begin{equation}
\rho_s \kappa\hat{\bf z}\times ({\bf v}_V-{\bf v}_s)+
D^\prime \hat{\bf z}\times({\bf
v}_n-{\bf v}_V)+D({\bf v}_n-{\bf v}_V) =0 ,
\label{1}
\end{equation}
where the first two terms represent the transverse force on the
vortex, while the parameter $D$ is responsible for the dissipative
friction ($\rho_s$ and ${\bf v}_s$ are superfluid density and
velocity,   ${\bf v}_n$ is the normal or  heat-bath velocity). The
measured ratio of two reactive parameters $d_\perp =D^\prime
/\kappa\rho_s$
\cite{Bevan1995}  reproduces at low $T$  the  result
$d_\perp\approx
0$ observed in the limit  $T\ll T_c$  \cite{Packard}. When $T$
increases,
$d_\perp$
first
becomes negative, then after reaching the minimum at $T\sim 0.4 T_c$
it increases, changes sign
and smoothly approaches
$d_\perp(T_c)=1$. The Eq.(1) of TAN \cite{Thouless}
suggests $d_\perp(T)=\rho_n(T)/\rho$, while the rest
of TAN paper implies that $d_\perp(T)=0$ at all $T$.
Both results are in disagreement with experiment and with correct
theory.

The reason is that  the formalism of  TAN does not
incorporate
the kinetic properties of fermions  localized in the vortex core
and  interacting with heat-bath fermions. This kinetics, determined
by the level spacing
$\omega_0$  and the life-time $\tau$ of the
core fermions
\cite{KopninCoAuthors,KopninVolovik1995,vanOtterlo1995,Stone}, leads to
\begin{eqnarray}
D\approx \kappa C_0~\tanh{\Delta(T)\over 2T} {\omega_0\tau  \over
1+\omega_0^2\tau^2};\\  D^\prime\approx \kappa \Bigl[
C_0-\rho_n(T)-{\omega_0^2\tau^2  \over 1+
\omega_0^2\tau^2}~C_0~\tanh{\Delta(T)\over 2T}~\Bigr] .\label{2}
\end{eqnarray}
The friction parameter $D$ is completely determined by the core
fermions:  the experimental bell
shape of  $d_\parallel(T)=D/\kappa\rho_s$ in Ref.\cite{Bevan1995}
follows  the $T$-dependence of $\omega_0\tau$ in the Eq.(2) with
$\omega_0\tau\gg 1$  at $T\ll T_c$
and $\omega_0\tau\ll 1$ close to $T_c$. The negative sign of
$d_\perp(T)$  observed by \cite{Bevan1995} at low $T$  is produced by
dominating
contribution of Iordanskii force, $D^\prime\approx -\kappa\rho_n(T)$
at $T\ll T_c$,
while
the core fermions with the $T$-independent spectral-flow parameter
$C_0=mp_F^3/3\pi^2$ are responsible for the observed upturn and
change of sign  of
$d_\perp(T)$ at $T>0.5~T_c$  \cite{KopninVolovik1995}.

The formalism used by TAN \cite{Thouless} and that in
\cite{KopninCoAuthors,KopninVolovik1995,vanOtterlo1995,Stone} lead to
different  results
due to the
effect similar to the axial anomaly in quantum field theory. Since the
core fermions  are
nearly gapless one should be extremely careful in which order to take
different limits. However within their theory TAN cannot resolve
between two different  regions
of
the kinetic parameter, $\omega_0\tau\ll 1$ and  $\omega_0\tau\gg 1$.
If $\omega_0\tau\ll 1$ the spectral flow of "chiral" core fermions
leads to an extra force on a vortex, which almost cancels the Magnus
force in superfluid/superconducting systems, where an approximate
particle-hole symmetry leads to $\rho-C_0\ll\rho$ \cite{Volovik1993}.
At low $T$ in many  (but not all)
systems
$\omega_0\tau \gg 1$: in this regime the discrete character of the
core  spectrum becomes
relevant, the spectral flow is suppressed and the full-size  Magnus
force  discussed by
TAN is restored. A similar effect of gapless fermions is responsible
for linear and  angular
momentum
paradoxes in the gapless $^3$He-A. An intrinsic dynamical  angular
momentum is a   small
fraction
$(\rho-C_0)/\rho$ of the value obtained in the similar density-matrix
formalism  (see
Refs.\cite{Volovik1986,Gaitan}). The linear momentum paradox in
$^3$He-A is  directly
related
to the axial anomaly: The direct derivation of the  momentum exchange
from the  anomaly
equation
$\partial_\mu j^\mu \sim FF^*$ shows that the effective Magnus force
on a  continuous
vortex
in $^3$He-A is reduced by the factor $(\rho-C_0)/\rho$
\cite{Volovik1992}. For such continuous vortex, $\omega_0$ is very
small and this  reduction
ceases only at very low $T$ \cite{Kopnin1995}. Such anomaly is
apparently  missing in
\cite{Thouless}.

\end{document}